# First Principles Investigations of the Atomic, Electronic, and Thermoelectric Properties of Equilibrium and Strained Bi$_2$Se$_3$ & Bi$_2$Te$_3$, with van der Waals Interactions

Xin Luo, Michael B. Sullivan, and Su Ying Quek[1)]

Institute of High Performance Computing, Agency for Science, Technology and Research, 1 Fusionopolis Way, #16-16 Connexis, Singapore 138632

**Abstract**

Bi$_2$Se$_3$ and Bi$_2$Te$_3$ are layered compounds of technological importance, being excellent thermoelectric materials as well as topological insulators. We report density functional theory (DFT) calculations of the atomic, electronic and thermoelectric properties of strained bulk and thin film Bi$_2$Se$_3$ and Bi$_2$Te$_3$, focusing on an appropriate description of van der Waals (vdW) interactions. The calculations show that the van der Waals Density functional (vdW-DF) with Cooper's exchange (vdW-DF$^{C09}_X$) can reproduce closely the experimental interlayer distances in unstrained Bi$_2$Se$_3$ and Bi$_2$Te$_3$. Interestingly, we predict atomic structures that are in much better agreement with the experimentally determined structure from Nakajima than that obtained from Wyckoff, especially for Bi$_2$Se$_3$ where the difference in atomic structures *qualitatively* changes the electronic band structure. The band structure obtained using the Nakajima structure, and the vdW-DF$^{C09}_X$ optimized structure, are in much better agreement with previous reports of photoemission measurements, than that obtained using the Wyckoff structure. Using vdW-DF$^{C09}_X$ to fully optimize atomic structures of bulk and thin film Bi$_2$Se$_3$ and Bi$_2$Te$_3$ under different in-plane and uniaxial strains, we predict that the electronic band gap of both the bulk materials and thin films decreases with tensile in-plane strain and increases with compressive in-plane strain. We also predict, using the semiclassical Boltzmann approach, that the magnitude of the *n*-type Seebeck coefficient of Bi$_2$Te$_3$ can be increased by the compressive in-plane strain, while that of Bi$_2$Se$_3$ can be increased with tensile in-plane strain. Further, the in-plane power factor of *n*-doped Bi$_2$Se$_3$ can be increased with compressive uniaxial strain, while that of *n*-doped Bi$_2$Te$_3$ can be increased by compressive in-plane strain. Strain engineering thus provides a

---

[1)] Electronic mail: queksy@ihpc.a-star.edu.sg

direct method to control the electronic and thermoelectric properties in these thermoelectric topological insulator materials.

I. INTRODUCTION

$Bi_2Se_3$ and $Bi_2Te_3$ are members of the $(Bi, Sb)_2(Te, Se)_3$ family of traditional thermoelectric materials - they can directly convert waste heat to electricity without any moving parts. These bulk thermoelectric materials were discovered to have large Seebeck coefficients half a century ago and are now widely used in thermoelectric refrigeration.[1-3] In recent years, there has been a surge of renewed interest in these thermoelectric materials – it was predicted and later, experimentally demonstrated,[4-7] that these materials constitute an exotic class of condensed matter, called topological insulators.[8-10] The topological insulators are distinguished by the existence of metallic spin-helical surface states, which are robust against the presence of nonmagnetic impurities or disorder.[11, 12] These surface states have potential applications in spintronics,[9, 13] quantum computation,[8, 10, 14] and thermoelectric energy conversion.[15] Importantly, these applications require a better fundamental understanding of the atomic and electronic structure of $Bi_2Se_3$ and $Bi_2Te_3$ when interfaced with other materials.

It is well known that $Bi_2Se_3$ and $Bi_2Te_3$ belong to the tetradymite-type crystal with a rhombohedral structure (point group *R-3m*). In the rhombohedral unit cell (Fig. 1a), there are three Se (Te) atoms that can be classified into two inequivalent types. We label these inequivalent atoms as $Se_1(Te_1)$ (two of them in one unit cell) and $Se_2(Te_2)$. The Bi atoms are equivalent. The $Bi_2Se_3$ and $Bi_2Te_3$ structure is also often alternatively described in the hexagonal representation with a unit cell of 15 atoms, as shown in Fig.1b. Within this representation, it is clear that $Bi_2Se_3$ and $Bi_2Te_3$ have layered structures. Each $Se_1(Te_1)$-Bi-$Se_2(Te_2)$-Bi-$Se_1(Te_1)$ forms a so-called quintuple layer (QL), which is a slab with five atomic layers. The QLs are stacked along the c-axis with the weak van der Waals interactions between neighboring QLs. The van der Waals (vdW) interaction is relatively weak, but it can play a dominant role in interactions between atoms or layers separated by empty space (so-called sparse matter). This interaction results exclusively from long-range correlations, which are absent from standard local and gradient-corrected DFT functionals.[16, 17] Much significant advancement has since been made that enables the treatment of vdW interactions within DFT. In addition to the method of dispersion-correction as an add-on to DFT,[18] the recently developed vdW-DF functional [19, 20] incorporates the long-range dispersion effects as a perturbation to the local density approximation correlation term, and this method has been applied successfully in diverse material systems.[17] The choice of exchange functional is also important – the standard functional used within vdW-DF, revPBE[21] typically gives vdW bond lengths that are a few per cent too large.[22] Most recently, Cooper developed an exchange functional that reduces the short-range repulsion term in revPBE.[23]

Although extensive electronic structure calculations have been performed for $Bi_2Se_3$ and $Bi_2Te_3$,[13, 24-31] most of them are calculated with experimental structures without full relaxation, or in the slab calculations, with only the top four layers of atoms in the top QL allowed to relax, fixing the inter-QL distance.[13, 32]

Yet, inter-QL van der Waals interactions are essential for predicting atomic and electronic structures of $Bi_2Se_3$ and $Bi_2Te_3$, when interfaced or intercalated with other materials. Indeed, much of the current interest in these materials involves interfacing them with other materials, and recent experiments indicate that depositing Ag on $Bi_2Se_3$ results in Ag intercalation between QLs.[11, 33] Furthermore, van der Waals interactions are required for accurate predictions of atomic structures of $Bi_2Se_3$ and $Bi_2Te_3$ under strain, which can directly influence topological properties.[28, 34, 35] On the other hand, previous theoretical calculations found that pressure and uniaxial stress can greatly influence the thermoelectric properties of $Sb_2Te_3$,[36] stress also plays an important role in the formation of defects in these thermoelectric materials.[37, 38] Recent experiments and molecular simulations show that the lattice thermal conductivity of thermoelectric materials will be affected by different strain conditions[39, 40]. How important is the van der Waals interaction in strain engineering, and how do they affect the thermoelectric properties of these materials?

In this work, we first explore the applicability of different exchange-correlation functionals, including those with vdW corrections, on predicting atomic structures of $Bi_2Se_3$ and $Bi_2Te_3$. Next, using an appropriate vdW functional, we fully optimize the atomic structures of strained bulk and thin film $Bi_2Se_3$ and $Bi_2Te_3$. Based on these optimized reference structures, the effect of strain on atomic, electronic and thermoelectric properties are reported, and the importance of vdW interactions is elucidated by comparing the results with those obtained using structures optimized with the PBE functional (including spin-orbit interactions). We also explore the effects of vdW interactions and spin-orbit interactions on the bulk moduli and phonon frequencies in the unstrained bulk systems.

## II. COMPUTATIONAL DETAILS

Except for the thermoelectric transport properties (addressed below), all our calculations are performed using the plane wave density functional theory (DFT) code, Quantum-Espresso (QE).[34] The norm-conserving pseudopotentials are generated using the Rappe-Rabe-Kaxiras-Joannopoulos (RRKJ) approach. For structural relaxation, the plane-wave kinetic energy cutoff is set to 56 Ry and the Brillouin zone is sampled with a 9x9x1 Monkhorst-Pack mesh - using a higher plane-wave cutoff of 70 Ry and a 13x13x1 k-point mesh changes the lattice constants and internal coordinates by less than 0.2%, with essentially no difference in resulting band structure. Phonon frequencies in the bulk are computed using a dense k-point mesh of 13x13x13. On the other hand, a plane-wave cutoff of 40 Ry is found to be sufficient (compared to 56 Ry) for calculations of band structure and bulk moduli. A vacuum thickness of 16Å is used in thin film slab calculations (converged relative to a vacuum thickness of 20 Å). In the self-consistent calculation, the convergence threshold for energy is set to $10^{-9}$ eV. All the internal atomic coordinates and lattice constants are relaxed, until the maximum component of Hellmann-Feynman force acting on each ion is less than 0.003 eV/Å. The spin-orbit coupling (SOC) effect, important for the heavy elements considered here, is treated self-consistently in fully relativistic pseudopotentials for the valence electrons.[41] In order to investigate the importance of SOC and vdW interactions for different physical properties, we have performed a detailed investigation with different exchange-correlation functionals. However, the main physical insights are obtained using atomic structures optimized using the vdW-DF functional[19, 20, 42] with Cooper's exchange,[23] and electronic structure calculated using the PBE[43] functional with SOC.

Many thermoelectric calculations[44-46] of similar materials are based on the WIEN2K package[47] with the semi-classical Boltzmann transport method in the relaxation time approximation[48]. Therefore, to compare our results with the literature, we utilize WIEN2K with BoltzTrap[45] for calculating thermoelectric transport properties of our previously relaxed structures, with the same basis functions as in the reported literature,[44, 45] and using the PBE[43] functional with SOC [49] as implemented in WIEN2K. The calculation of transport properties requires a very dense k-grid; here, a non-shifted mesh with 56000 k points (4960 in the IBZ) is used, which is found to be converged as compared to a denser sampling with 70000 k-points. Within BoltzTrap, the relaxation time τ is assumed to be a constant with respect to the wave vector k and energy around the Fermi level, and the effect of doping is introduced by the rigid band approximation. Within the relaxation time approximation, the Seebeck coefficient $S$ can be obtained directly from the electronic structure without any adjustable parameters.

## III. RESULTS AND DISCUSSIONS

### A. Structural Properties of Equilibrium $Bi_2Se_3$ and $Bi_2Te_3$

Table 1 shows the fully optimized lattice constants and internal coordinates obtained using different functionals. The different functionals give consistent internal coordinates, but the lattice parameters, especially that related to the interlayer distance $d_{eqm}$ (vertical distance between Se (or Te) atoms in adjacent QLs), vary significantly with different functionals. Focusing first on the LDA, PBE and revPBE functionals, we find that LDA[50] results are closest to experiment, with LDA underestimating $d_{eqm}$ by 6.7% and 5.6% for $Bi_2Se_3$ and $Bi_2Te_3$ respectively, and PBE[43] overestimating $d_{eqm}$ by 38.6% and 15.8% for $Bi_2Se_3$ and $Bi_2Te_3$ respectively. The revPBE[21] functional predicts an even larger $d_{eqm}$ of > 4.48Å. Calculations with the vdW-DF correlation functional together with revPBE exchange lead to improved, smaller $d_{eqm}$ compared to the revPBE functional, however, the predicted $d_{eqm}$ is still overestimated by more than 20%. This error is much larger than that reported for other vdW-bonded systems.[22] In contrast, the semi-empirical vdW correction based on Grimme's scheme (PBE-D)[18, 51, 52] does help, predicting $d_{eqm}$ within 3.4% and 8% of experiment for $Bi_2Se_3$ and $Bi_2Te_3$, respectively. On the other hand, using the vdW-DF functional [19, 53] with the most recently developed Cooper's exchange (vdW-DF$^{C09}{}_X$),[23] we can obtain better agreement with experiment, with predicted $d_{eqm}$ values of 2.58Å for $Bi_2Se_3$ and 2.582Å for $Bi_2Te_3$, which are within 0.1 % and 1.1% of experiment. We next consider the role of spin-orbit coupling (SOC) in relaxing atomic structures. We find that SOC tends to reduce $d_{eqm}$ in all cases; as a result, adding SOC increases the error for LDA, but reduces the error for PBE and PBE-D, thus giving $d_{eqm}$ values within 1.5% of experiment for PBE-D+SOC. As the vdW-DF functional has not been implemented with SOC, we did not check the effect of SOC on these results. However, since vdW-DF$^{C09}_X$ can already give excellent agreement with experiment, it is still unclear if SOC is truly important for structural optimization. We note that the very reasonable results of PBE-D and PBE-D+SOC are quite remarkable given that the correction is semi-empirical. However, in the following, we shall base most of our conclusions on structures optimized using the vdW-DF$^{C09}_X$ functional.

Since the interlayer vdW interactions are important, we focus on these interactions by computing, for simplicity, the inter-QL binding energy as a function of interlayer distance, for 2QL Bi$_2$Se$_3$ and Bi$_2$Te$_3$ thin films (i.e. films with two QLs each containing five atomic layers), where the internal atoms within each QL are fixed to their experimental atomic positions from the bulk (Fig. 2). Although the inter-QL distance for the 2QL film may be slightly different from the bulk, it has been shown that the vdW interactions are dominated by those between nearest neighbour layers in other layered materials such as multi-layer graphene and MoS$_2$.[54, 55] The close correspondence (with deviation of less than 5% for all functionals) between the equilibrium inter-QL distances calculated for the bulk and for 2 QLs (Table 1) further supports this assumption. From Fig. 2, we see that revPBE and revPBE+SOC give rise to repulsive interactions between the 2QLs. Furthermore, the curvature of the energy versus distance curve is quite different for different functionals. Although SOC does not change $d_{eqm}$ significantly, it does affect the interlayer force constants, which are given by the second derivative of the energy as a function of distance. We also note that the binding energies obtained using vdW-DF$^{C09}_X$ or PBE-D are consistent with recent estimates obtained by rescaling results from the VV10 functional[56] according to quantitative random-phase approximation (RPA) calculations.[57] These binding energies are larger for Bi$_2$Te$_3$ than for Bi$_2$Se$_3$, a finding consistent with the larger atomic size of Te compared to Se.

Interestingly, we find that our first-principles-predicted structures using both PBE-D and vdW-DF$^{C09}_X$ functionals are more consistent with the experimental structures reported by Nakajima,[58] rather than that from Wyckoff (Table 1).[59] Both experimental structures have been widely used in the theoretical literature, but importantly, the internal atomic coordinates are different in these two structures (Table 1), especially for the Bi$_2$Se$_3$, where the interlayer distance $d_{eqm}$ is quite different (2.579 Å in Nakajima's structure and 2.253 Å in Wyckoff's structure). It is noted that the crystal structure from Wyckoff is obtained from an early electron diffraction study,[60] while structures of Nakajima are obtained from X-ray diffraction powder analysis. As the interaction of electrons with matter are about 10,000 times stronger than that of X-rays, multiple dynamical scattering will influence the intensity of electron diffraction patterns, thus making the structure determination from electron diffraction more difficult and less reliable than that from x-ray diffraction.[58]

## B. Electronic Properties of Equilibrium $Bi_2Se_3$ and $Bi_2Te_3$

Using PBE+SOC, we compute the band structures of the Nakajima and Wyckoff atomic structures, and the vdW-DF$^{C09}_X$ optimized structures. We find that the band structures for the vdW-DF$^{C09}_X$ optimized structures match very well to those of the Nakajima structures, as we expect from the above discussion. The band structures of the Wyckoff and Nakajima structures are very similar in the case of $Bi_2Te_3$ (Fig. 3b), but are *qualitatively* different for $Bi_2Se_3$ (Fig. 3a). Focusing now on these differences for $Bi_2Se_3$, we note that the lowest conduction band along the $\Gamma$ - Z - F high symmetry direction is more rippled in the Wyckoff structure than in the Nakajima and vdW-DF$^{C09}_X$ optimized structures (Fig. 3a). Furthermore, the direct gap at F is 0.40 eV smaller in the Wyckoff $Bi_2Se_3$ structure than in the Nakajima structure, while the band gap at $\Gamma$ is 0.20 eV smaller in the Nakajima $Bi_2Se_3$ structure than in the Wyckoff structure. (We note that the calculated bandstructures of $Bi_2Se_3$ based on the PBE-D+SOC relaxed structure and that of vdW-DF$^{C09}_X$ optimized structures are almost the same; not shown here.) Comparing the computed band structure with that measured in recent angle resolved photoelectron spectroscopy (ARPES) experiments,[61] we find that the band structure for the Nakajima structure is closer to experiment than that for the Wyckoff structure, thus providing direct experimental evidence favouring the coordinates from Nakajima and our optimization procedure. Specifically, there are two key qualitative features indicating that the ARPES measurements agree better with the band structure for the Nakajima structure. Focusing on the band structure along $\Gamma$ - Z, close to the Fermi level, we see: 1. the highest occupied valence band is flatter in the Nakajima structure, similar to that from ARPES; 2. the energy separation between the highest occupied valence band and the next highest occupied band is about 0.75 eV for the Nakajima structure, in contrast to 0.3 eV for the Wyckoff structure, and similar to that from ARPES. It is thus clear that agreement is much better for the Nakajima structure than for the Wyckoff structure. Interestingly, the authors of the ARPES paper[61] had commented that the measured energy bands in the $\Gamma$ - Z direction were significantly flatter than that predicted by DFT; our calculations strongly suggest that the reason for this discrepancy was that the Wyckoff structure was used for their calculations, instead of the Nakajima structure. Since the main difference between the Nakajima and Wyckoff $Bi_2Se_3$ structures is the larger interlayer separation in the former, these results underscore the importance of a careful treatment of vdW interactions for prediction of band structures.

We note that although DFT in many cases gives accurate qualitative predictions of band structure, the DFT Kohn-Sham values cannot quantitatively predict quasiparticle band structures.[62] In this case, the calculated band gaps of $Bi_2Se_3$ and $Bi_2Te_3$ (using the vdW-$DF^{C09}_X$ optimized structures) are 0.30 eV and 0.10 eV, slightly smaller than the experimentally measured values of 0.35 eV for $Bi_2Se_3$[63] and 0.17 eV for $Bi_2Te_3$,[6] but consistent with the other DFT results.[13, 26]

Although our calculations suggest that spin-orbit coupling (SOC) is not crucial for predicting atomic structures, SOC has an important effect on the band structure. To illustrate this, we compare the band structures in Fig. 3a-b with those computed without SOC (Fig. 3c-d), using the vdW-$DF^{C09}_X$ optimized structures. The band structures with and without SOC are quite different, consistent with large SOC-induced band inversion at Γ. Furthermore, we note that the band structures computed by the vdW-$DF^{C09}_X$ functional and PBE are quite similar, suggesting that the vdW-$DF^{C09}_X$ functional can produce band structures that are consistent with PBE, and that vdW interactions have minimal effect on the band structure. This justifies our neglect of vdW effects in the band structure calculations.

Recent interest in the topological metallic surface states in $Bi_2Se_3$ and $Bi_2Te_3$ slabs focus on the Dirac cone feature in the band structure at the Γ point.[13, 30, 64-66] At very thin slab thickness, the interaction between the surface states at opposite interfaces opens the band gap at Γ, and this gap will gradually decay to zero as the slab thickness is increased. Previous PBE (with SOC) calculations on the unrelaxed slabs derived from Wyckoff structures found that the band gap at Γ begins to close by 3 and 4 QL for $Bi_2Se_3$ and $Bi_2Te_3$, respectively[65]. To address the question of whether the atomic structure influences these predictions, we fully optimized slab structures 1-5 QL thick using the vdW-$DF^{C09}_x$ functionals; the fully optimized slab structures are similar to those directly derived from Nakajima structures. Based on the optimized slab structures, we calculated the band gap at Γ with PBE+SOC, as shown in Fig.4. Our results are close to that using the Wyckoff structures, except that the band gap closes only at 4 QL instead of 3 QL for $Bi_2Se_3$. We note that compared to experiment and previous GW calculations,[64] the predicted gaps are too small. It is unclear if using the Nakajima instead of the Wyckoff structures would affect the computed GW gaps; however, such calculations are beyond the scope of the current work.

### C. Bulk Moduli and Phonon Frequencies of Equilibrium $Bi_2Se_3$ and $Bi_2Te_3$

We further predict the bulk moduli and phonon frequencies of equilibrium $Bi_2Se_3$ and $Bi_2Te_3$ with vdW-DF$^{C09}_X$ and other functionals, and compare our predictions with available experimental literature. The bulk modulus $B$ and its pressure derivative $B'$ can be obtained by computing the changes in total energy of $Bi_2Se_3$ and $Bi_2Te_3$ with hydrostatic pressure, and fitting the resulting energy-volume curves using the Birch-Murnaghan equation of state:

$$E(\eta) = E_0 + \frac{9BV_0}{16}\left[\left(\eta^2 - 1\right)^3 B' + \left(\eta^2 - 1\right)^2\left(6 - 4\eta^2\right)\right] \quad (1)$$

Where $\eta = (V_0/V)^{1/3}$, $V_0$ is the equilibrium volume of the fully relaxed structure. The bulk moduli thus obtained are reported in Table 2. In general, the values of $B$ predicted for both $Bi_2Se_3$ and $Bi_2Te_3$ are similar (~30-40GPa). The experimental reports for $B$ differ, especially for $Bi_2Se_3$ where an exceptionally high value of 53GPa is obtained by Vilaplana et al.[67]. However, the remaining experimental values are in reasonable agreement with our predicted values. Due to the different experimental values, we are unable to draw a conclusion of the importance of van der Waals interactions or spin-orbit coupling in determining the bulk modulus.

More insights can be obtained by comparing computed zone center phonon frequencies with experiment. As can be shown by group theory, the irreducible representations of the zone center phonon modes in $Bi_2Te_3$ and $Bi_2Se_3$ are $\Gamma = 2A_{1g} + 3A_{2u} + 2E_g + 3E_u$, among which $2A_{1g}$ and $2E_g$ are Raman-active; $3A_{2u}$ and $3E_u$ are Infrared-active, the three acoustic modes are composed by one $A_{2u}$ and two degenerate $E_u$ modes. For the Raman active modes, $E_g$ describes the shear mode of in-plane atomic vibrations and $A_{1g}$ describes the breathing mode of out-of-plane atomic vibrations. Except for results obtained with the vdW-DF functional, all phonon frequencies are computed within density-functional perturbation theory (DFPT) as introduced by Lazzeri and Mauri[68] in Quantum Espresso. Since the phonon frequencies cannot be computed with vdW-DF in Quantum Espresso, we adopt the force-constant approach[69, 70] to compute the phonon frequencies with the vdW-DF$^{C09}_X$ functional. In this method, we displace each atom in the primitive cell from its equilibrium position in the $x$, $y$, and $z$ directions by a distance of 0.015 Å, and calculate the forces acting on each atom using the Hellmann-Feynman theorem. Subsequently, the interatomic force-constant matrix is evaluated using a central finite-difference scheme. (We have checked that within LDA, the phonon frequencies as calculated with the force-constant approach are essentially the same as those obtained from the DFPT method.)

First, comparing LDA and PBE results with and without spin-orbit coupling (SOC) with experiment (Table 3), we note that in all cases, SOC reduces the phonon frequencies. In general, inclusion of SOC then leads to better agreement with experiment, as previously observed by Cheng et al.[71]. We note that the importance of SOC in determining force constants is in contrast to our earlier observation that SOC was not important for structural relaxation. The difference lies in the fact that force constants are related to the rate of change of force with atomic displacements, whereas structural relaxation is related only to the coordinates for the minima of the potential energy surface. This can be illustrated in the simplified picture in Fig. 2 where the minima of the energy-distance curves are similar with and without SOC, but the curvatures of the curves are quite different. Next, although the PBE atomic structure is significantly different from experiment compared to the PBE-D atomic structure, we find that the PBE-D frequencies are quite similar, and in fact, the PBE-D frequencies are in many cases slightly farther from the experimental values. The PBE-D frequencies are quite similar to those obtained with the vdW-DF$^{C09}_X$ functional. However, PBE-D+SOC frequencies give the best match with experiment, suggesting that PBE-D+SOC can be used to make predictions on frequencies.

### D.  Atomic and Electronic Properties of Strained Bi$_2$Se$_3$ and Bi$_2$Te$_3$

Strain engineering is a mature technique for controlling the electronic properties of nanoscale semiconductors in industry - mechanical strain can be imposed by microelectromechanical systems (MEMS) or by epitaxial growth of thin films. To study the effect of strain on the atomic structures and electronic properties of $Bi_2Se_3$ and $Bi_2Te_3$, we imposed different in-plane and uniaxial strains to the bulk material. The strain is defined by $\eta = (\alpha_{strain}/\alpha_0 - 1) \times 100\%$, where $\alpha_{strain}$ is the lattice constant of the strained state, and $\alpha_0$ is the optimized lattice constant of the unstrained bulk material. $\alpha$ is the in-plane lattice constant and out-of-plane lattice constant for in-plane strain and uniaxial strain, respectively. In our calculation, a uniform in-plane strain from -3% to 3% is applied to $Bi_2Se_3$ and $Bi_2Te_3$ bulk and 2QL thin films. For the bulk material, the effect of uniaxial strain from -6% to 6% is also investigated. To predict the strained structures in the bulk, we start by constraining the in-plane (or out-of-plane) lattice constants to the strained state, and then relaxing the lattice constant along the c-axis (or uniformly in-plane), as well as the internal coordinates, with the vdW-DF$^{C09}_X$ functional. Once we obtain the optimized atomic structures, we compute the electronic properties using PBE+SOC.

With the in-plane strain changing from 3% to -3%, we find that the out-of-plane lattice constants of the relaxed bulk structures increase approximately linearly, with a slope of 0.342 Å (1.2% of out-of-plane lattice constant) and 0.353 Å (1.17% of out-of-plane lattice constant) per unit decrease in percentage in-plane strain, for $Bi_2Se_3$ and $Bi_2Te_3$, respectively. For uniaxial strain changing from 6% to -6%, a similar linear relationship was found, with the slope of 0.008 Å (0.19% of in-plane lattice constant) and 0.012 Å (0.27% of in-plane lattice constant) per unit decrease in percentage uniaxial strain for $Bi_2Se_3$ and $Bi_2Te_3$, respectively. These results show that the out-of-plane lattice constant is strongly coupled with the in-plane lattice constant when the structures are optimized under different strain. The different percentage changes suggest strong anisotropy in the elastic properties of these layered materials. Although the overall band structures of the strained systems are similar to those of the unstrained ones, the band structure near the Fermi level is influenced by the applied strains, resulting in significant changes in the band gaps. Fig. 5a shows the evolution of band gap as a function of the in-plane strain for bulk $Bi_2Se_3$ and $Bi_2Te_3$. Focusing on the vdW-DF$^{C09}_X$ relaxed structures, we note that the energy gap increased from 0.07eV to 0.16eV for bulk $Bi_2Te_3$, when the in-plane strain changed from 3 % to -3 %. For vdW-DF$^{C09}_X$ optimized bulk $Bi_2Se_3$, the band gap in general increases from extensive to compressive strain (from 0.25eV (3.2 % strain) to 0.33eV (-1.8 % strain)). However, further increase in compressive strain to -2.8% reduces the band gap (here, the k-point for the valence band maximum shifts from its original location near the Z symmetry point to the shoulder of the M shape band near the Gamma high symmetry point). To illustrate the importance of vdW interactions, we relaxed the structures with both vdW-DF$^{C09}_X$ and PBE+SOC functionals, and then calculated the electronic structures with PBE+SOC. The band gaps for the PBE+SOC optimized bulk Bi$_2$Te$_3$ are similar to those for the vdW-DF$^{C09}_X$ relaxed structures. However, the band gaps obtained for PBE+SOC optimized bulk Bi$_2$Se$_3$ are significantly different. These results are consistent with the fact that PBE+SOC overestimates $d_{eqm}$ by 28% in bulk Bi$_2$Se$_3$, but by only 4.7 % in bulk Bi$_2$Te$_3$.

Moving now to the 2QL thin films, optimized using vdW-DF$^{C09}_X$, we note that the optimized in-plane lattice constants are smaller than the bulk - about 0.3% and 1.0% smaller for Bi$_2$Te$_3$ and Bi$_2$Se$_3$, respectively. However, the equilibrium inter-QL distances ($d_{eqm}$ is 2.589 Å for Bi$_2$Se$_3$ and 2.595 Å for Bi$_2$Te$_3$ in the fully relaxed 2QL thin films) are similar to those obtained in Fig. 2. The band gap for 2QL films is smaller than that for the bulk, because the band gap in 2QL films is determined by interactions between metallic surface states on both sides of the film. As observed for the bulk, the band gap for the 2QL films increases with compressive strain, and decreases with tensile strain, with the exception of Bi$_2$Te$_3$ thin films, where there is a very small decrease in band gap at -3% compressive strain (Fig. 5b). Further, the Bi$_2$Se$_3$ and Bi$_2$Te$_3$ thin films become metallic when the tensile strain is larger than 3% and 1%, respectively. We note that in this case, using the PBE+SOC relaxed structures results in band gaps that are about twice as large or more, although the general trend of how the band gap evolves with strain are consistent with those for the vdW-DF$^{C09}_X$ optimized structures.

It should be noted that in both the thin films and bulk material, the band gap is indirect. The direct band gap at the Γ point is an important issue for topological insulators, because of the Dirac cones at Γ for the thin films. Fig. 5c and 5d show the band gaps at Γ for both the bulk and the 2QL thin film. We note that Bi$_2$Se$_3$ has a smaller direct band gap at Γ than Bi$_2$Te$_3$, but has a larger indirect band gap; this explains why Bi$_2$Se$_3$ is more widely studied for its potential applications as a topological insulator, where it is important to distinguish metallic surface state carriers from intrinsic bulk carriers. Except for Bi$_2$Se$_3$ 2QL films, the direct band gap at Γ tends to decrease from extensive to compressive strain. Predictions using PBE+SOC relaxed structures result in similar trends for Bi$_2$Te$_3$ but not for Bi$_2$Se$_3$.

We note that Young et al. has investigated the evolution of the topological phase of bulk Bi$_2$Se$_3$ under mechanical strain,[28] using regression fits to obtain band gap stress and stiffness tensors (the linear and quadratic coefficients relating the Γ point band gap to strain). From these tensors, it was predicted that the topological phase transition will occur at 6.4% uniaxial strain in the out-of-plane direction, relative to the experimental structure. Here, we apply uniaxial strains on bulk Bi$_2$Se$_3$ and Bi$_2$Te$_3$, in each case fully optimizing the internal coordinates and in-plane lattice constants using the vdW-DF$^{C09}_X$ functional. In contrast to the case of in-plane strain, which affects the out-of-plane lattice constant significantly, out-of-plane uniaxial strain has much less effect on the in-plane lattice constant. In general, the gaps decrease with increasing tensile uniaxial strain (Fig. 6a), and there is no clear quadratic or linear relation of the uniaxial strain at Γ point band gap in Bi$_2$Se$_3$. Our calculations also predict a topological phase transition (closing of Γ point band gap) at 6% uniaxial strain for Bi$_2$Se$_3$ (Fig. 6b), which interestingly, is approximately consistent with the prediction by Young et al. Finally, we note that the Γ point band gap is more sensitive to uniaxial strain than the indirect band gap.

To assess the practical feasibility of strain engineering, we compare the strain energy (the energy differences between the strained states and their corresponding unstrained counterparts) with the vdW interaction energy. The strain energy is hundredths of eV/unit cell, and even at 3% in-plane strain or 6% uniaxial strain, is still less than 0.1 eV/unit cell. This is quite small compared to the vdW inter-QL interaction energy of about 0.20-0.25 eV/unit cell (Fig. 2), indicating that these layered compounds are likely to be able to withstand such strains without surface exfoliation, the prospects of using strain (eg. via MEMS) to engineer their band gaps are promising.

### E. Thermoelectric properties

As mentioned before, Bi$_2$Te$_3$ and Bi$_2$Se$_3$ are traditional thermoelectric materials that can generate electricity from waste heat once a temperature gradient exists.[46, 72-74] The thermoelectric performance is quantified by the figure of merit, *ZT*, where *T* is the temperature and *Z* is defined by[3]

$$Z = \frac{S^2 \sigma}{(\kappa_e + k_L)} \quad (2)$$

$S$ is the Seebeck coefficient, $\sigma$ the electronic conductivity, and $\kappa_e$ and $k_L$ are the electronic and lattice thermal conductivities, respectively. $S^2\sigma$ is also known as the power factor. Superlattice materials based on $Bi_2Te_3$ have resulted in high $ZT$ values > 1.[3] A higher $ZT$ can be obtained by increasing the power factor and decreasing the thermal conductivity.[75] In the following, we shall focus on analyzing the effect of strain on the power factor and the Seebeck coefficient, which can be computed within the linear response regime in the semi-classical Boltzmann framework,[45] as:

$$\sigma \equiv q^2 L_0 \qquad (3)$$

$$S = \frac{k_B}{q}\frac{L_1}{L_0} \qquad (4)$$

with

$$L_j = \int_{-\infty}^{\infty} -\frac{\partial f_0}{\partial E} D(E) v^2 \tau \left(\frac{E-\mu}{k_B T}\right)^j dE$$

where $q$ is the charge of carriers, $f_0$ is the Fermi distribution function of electrons, $v$ is the Fermi velocity, $\tau$ is the relaxation time, $\mu$ is the chemical potential, and $D(E)$ is the density of states. Using a constant relaxation time approximation, the Seebeck coefficient can be completely determined from the band structure.

We note that several scholars have done some pioneering thermoelectric studies on related materials, using PBE+SOC optimized structures, or experimental structures.[36, 44] It was shown that the semiclassical Boltzmann transport method within the relaxation time approximation can predict thermoelectric properties of unstrained $Bi_2Te_3$, in good agreement with experiment.[44] Furthermore, it was predicted that strain engineering can increase the power factor in $Sb_2Te_3$.[36, 38] More recently, studies on Bi and Sb tellurides show that the lattice constants and volume expansion have an important influence on the temperature behavior of Seebeck coefficient,[76] and strain can also play an important role in the anisotropy of electrical conductivity and Seebeck coefficient.[77] Since most of these studies are carried out in Wien2k using the semiclassical

Boltzmann transport method,[45] for the sake of comparison, we use the same method (and the same, tested parameters from the literature[44]) for calculation of thermoelectric properties.[78]

We study the effect of in-plane and out-of-plane strain on the in-plane Seebeck coefficient and power factor of bulk $Bi_2Se_3$ and $Bi_2Te_3$ under *n* and *p*-type doping, and at different temperatures, using the vdW-DF$^{C09}_X$ optimized structures. (The out-of-plane conductivity is very low, thus resulting in very small power factors.) We find that in-plane compressive strain can significantly improve the Seebeck coefficient and power factor of $Bi_2Te_3$ (by as much as two times), while effects on $Bi_2Se_3$ are less significant. Overall, $Bi_2Te_3$ has a power factor that is an order of magnitude larger than that of $Bi_2Se_3$, even in the unstrained state; therefore, these results are significant for further enhancing the *ZT* of this excellent thermoelectric material. In contrast, we find that uniaxial strain does not improve the power factor of $Bi_2Te_3$. For $Bi_2Se_3$, the power factor can be increased by compressive uniaxial strain in the case of *n*-doping, and by 6% uniaxial tensile strain in the case of *p*-doping.

Fig. 7 shows the calculated in-plane Seebeck coefficient *S* at 300K for both *n*-type and *p*-type doping as a function of carrier concentration. As the electron concentration increases, the Fermi level is shifted higher into the conduction band, resulting in less asymmetry between electrons and holes, therefore reducing the *n*-type Seebeck coefficient. Similar arguments can be made for *p*-doped systems. Focusing on the vdW-DF$^{C09}_X$ relaxed structures (Fig. 7a-b), we see that the Seebeck coefficient of *p*-doped $Bi_2Te_3$ can be increased by 2% tensile strain for hole concentrations of $10^{19}$-$10^{21}$ cm$^{-3}$, and by 2% compressive strain for hole concentrations of $10^{17}$-$10^{19}$ cm$^{-3}$. On the other hand, 2% compressive strain can increase the magnitude of *S* in *n*-doped $Bi_2Te_3$ for all dopant concentrations studied, while 2% tensile strain reduces the magnitude of *S*. This trend is consistent with the increase in band gap from tensile to compressive strain in $Bi_2Te_3$, as well as to the steeper slope of conductivity versus energy (as can be inferred from Fig. 9b). For *p*-doped $Bi_2Se_3$, the overall Seebeck coefficient is significantly higher than $Bi_2Te_3$ due to the steeper variation of the density of states in the valence bands,[46] and the effect of strain on the improvement of Seebeck coefficient for $Bi_2Se_3$ is less obvious. However, 2% tensile strain can increase the magnitude of *S* in *n*-doped $Bi_2Se_3$ for electron concentrations of $2\times10^{18}$ cm$^{-3}$—$10^{20}$ cm$^{-3}$. For comparison, we have also computed the thermoelectric properties of corresponding PBE+SOC optimized structures. The predicted thermoelectric properties can be *qualitatively* different, as shown in Fig. 7c-d, therefore underscoring the importance of vdW interactions in structural optimization and in predicting the effects of strain on electronic and thermoelectric properties. In the following, we focus on vdW-DF$^{C09}_X$ relaxed structures.

Fig. 8 depicts the temperature dependence of the Seebeck coefficient $S$ under different in-plane strains, with the carrier concentration fixed to $10^{19}$cm$^{-3}$. Fig. 8a shows that compressive strain can increase the Seebeck coefficient of *p*-doped $Bi_2Te_3$ when the temperature is higher than 350K, and the peak value can be shifted from 300K to 400K when a 3% compressive strain is applied. This result is very similar to the prediction of Ref. 76, in which the authors changed the lattice constants of $Bi_2Te_3$ to that of $Sb_2Te_3$ to model a compressive in-plane strain. Overall, the strain does not improve the *p*-type Seebeck coefficient of $Bi_2Se_3$, except for a small enhancement at temperatures higher than 500K (Fig. 8b). As noted before, the *p*-type Seebeck coefficient in general is larger in $Bi_2Se_3$ than in $Bi_2Te_3$, reaching about 400µV/K at 350K.

On the other hand, compressive strain increases the magnitude of the Seebeck coefficient in *n*-doped $Bi_2Te_3$ for all temperatures considered here, and the maximum of the Seebeck coefficient is shifted to higher temperature with larger compressive strain (Fig. 8c). Most notably, under 3% compressive strain, the magnitude of the Seebeck coefficient reaches a maximum of 305µV/K at 350K, roughly 45% higher than that of unstrained $Bi_2Te_3$ (210µV/K) at the same temperature. For *n*-doped $Bi_2Se_3$, compressive strains instead reduce the magnitude of the Seebeck coefficient, but tensile strain increases this magnitude for temperatures less than ~500K (Fig. 8d). Furthermore, the maximum magnitude of the Seebeck coefficient is shifted to lower temperatures at larger tensile strains. A maximum magnitude of 256 µV/K is achieved in the Seebeck coefficient of $Bi_2Se_3$ under 1% tensile strain, 13% larger than the maximum magnitude in the unstrained system. It is interesting to note that under 2% tensile strain, the maximum magnitude of Seebeck coefficient of *n*-type $Bi_2Se_3$ (252 µV/K) is at about 350K, the same temperature where *n*-type $Bi_2Te_3$ has the largest Seebeck coefficient under 3% compressive strain. Since the in-plane lattice constant of $Bi_2Se_3$ is about 5% smaller than that of $Bi_2Te_3$, it is possible that in a superlattice structure, the tensile strain in $Bi_2Se_3$ and compressive strain on $Bi_2Te_3$ result in a larger Seebeck coefficient. Together with the reduced lattice thermal conductivity in a superlattice structure, this may greatly improve the thermoelectric performance.

The electrical conductivity $\sigma$ can be readily calculated within the constant relaxation time approximation given an estimate for the relaxation time $\tau$. In practice, we can estimate $\tau$ by comparing our computed values of $\sigma/\tau$ and $S$ for unstrained systems, with experimentally[74, 79] measured values of $\sigma$ and $S$ at the same temperature (300 K), as described in Ref. 44. In this way, in-plane relaxation times of $2.2\times10^{-14}$ s and $2.7\times10^{-15}$ s are derived for $Bi_2Te_3$ and $Bi_2Se_3$, respectively. In Ref. 44, it was shown that this derived relaxation time for $Bi_2Te_3$ gives good agreement with experiment for different doping concentrations; here, we further assume that the relaxation time is independent of strain, as is also assumed by other authors[37].

The resulting conductivities and power factors (at 300K) are plotted in Fig. 9 for different in-plane strains (positive and negative carrier concentrations denote the $p$-type and $n$-type doping respectively). The conductivity of $Bi_2Te_3$ is 20 times larger than that of $Bi_2Se_3$; this is consistent with the physical picture that $Bi_2Te_3$ has a smaller band gap. Strain has a significantly larger effect on the conductivity of $Bi_2Te_3$ than $Bi_2Se_3$. Tensile strain increases the conductivity while compressive strain decreases the conductivity in $Bi_2Te_3$, for both $p$-type and $n$-type doping, an observation that is consistent with the increase in band gap from tensile to compressive strain. The Seebeck coefficient is related with the conductivity and the electron-hole asymmetry. When $E - \mu \gg k_B T$, equation (2) can be expressed in the Mott formulas[80]:

$$S = (\frac{\pi^2 k_B^2 T}{3e\sigma})\frac{d\sigma}{dE}|_{E=E_F} = \frac{\pi^2 k_B^2 T}{3e}\frac{d\ln\sigma}{dE}|_{E=E_F} \quad (5)$$

Therefore, we can understand the Seebeck coefficient from the energy derivative of the log scale conductivity. By comparing the conductivity and Seebeck coefficients in Fig. 9a and 9b, we find that qualitatively, the carrier concentration can be used as an approximate proxy for the energy scale – the Seebeck coefficient is larger when the slope of the conductivity curve is steeper. This is also consistent with the increased Seebeck coefficients with compressive strain for $Bi_2Te_3$. We further note that although the Seebeck coefficient of $Bi_2Se_3$ is in general larger than that of $Bi_2Te_3$, under compressive strain, the $n$-type Seebeck coefficient of $Bi_2Te_3$ can surpass that of $Bi_2Se_3$. This large enhancement in $n$-type Seebeck coefficient of $Bi_2Te_3$ with compressive strain is in agreement with Ref. 77. Finally, we computed the power factor for $Bi_2Se_3$ and $Bi_2Te_3$ under different strain states. The power factor for $Bi_2Te_3$ is one order of magnitude larger than that for $Bi_2Se_3$, due to the much larger conductivity in $Bi_2Te_3$. Although the compressive strains can reduce the electron conductivity of *n*-type $Bi_2Te_3$, it will also increase its Seebeck coefficient. The larger enhancement of the Seebeck coefficient makes it possible to compensate the reduction of conductivity, resulting in an enhancement of the *n*-type power factor for $Bi_2Te_3$, under compressive in-plane strain. This result is different from that in Ref. 77 - the discrepancy may come from the different methods for imposing strain (in Ref. 77, the strain is imposed by setting the in-plane and out-of-plane lattice constants of $Bi_2Te_3$ to that of $Sb_2Te_3$, without relaxing the internal coordinates).

Finally, we show in Fig. 10 the thermoelectric properties of $Bi_2Se_3$ and $Bi_2Te_3$ under uniaxial strain. In contrast to the case for in-plane strain, tensile uniaxial strain decreases the conductivity, and compressive uniaxial strain increases the conductivity for $Bi_2Te_3$, while tensile uniaxial strain increases Seebeck coefficient slightly, and compressive strain decreases Seebeck coefficient. Unlike the case of in-plane strain, these opposite effects on Seebeck coefficient and conductivity result in no enhancement of the power factor for $Bi_2Te_3$ under compressive in-plane strain. Uniaxial strain can be used to enhance the power factor of $Bi_2Se_3$, however: compressive strain enhances the *n*-type power factor because of an increased magnitude of Seebeck coefficient (the power factor more than doubles under compressive uniaxial strain of 4-6%, for doping concentrations between $10^{19}$ and $10^{20}$ cm$^{-3}$), while 2% compressive uniaxial strain can enhance the p-type power factor.

## IV. CONCLUSION

In summary, we have performed a comprehensive investigation of the effects of strain on the atomic, electronic and thermoelectric properties of $Bi_2Se_3$ and $Bi_2Te_3$, taking into account the vdW interlayer interactions using the vdW-DF$^{C09}_X$ functional. Our optimized, unstrained, structures are in much closer agreement with the experimental structure from Nakajima (determined by X-ray diffraction) than to that from Wyckoff (determined by electron diffraction).[58, 59] Importantly, the two experimental structures have qualitatively different band structures on $Bi_2Se_3$ – previously published photoemission results[61] on $Bi_2Se_3$ are in good qualitative agreement with the band structure of vdW-DF$^{C09}_X$ optimized structure and the Nakajima structure, but not the Wyckoff structure, and the Γ point band gap for the vdW-DF$^{C09}_X$ optimized $Bi_2Se_3$ thin films closes at 4QL instead of 3QL as previously reported using the Wyckoff structure. We predict that the band gaps of these materials increase from tensile to compressive in-plane strain, suggesting that compressive strain may be used to increase the bulk band gap of $Bi_2Se_3$, thus making it easier to distinguish the metallic topological surface state from intrinsic bulk carriers. We also confirm that a topological phase transition can occur in $Bi_2Se_3$ at 6% uniaxial strain, as predicted by Young et al.[28]. Strain can also be used to tune the thermoelectric properties of these materials: the $n$-type Seebeck coefficient of $Bi_2Te_3$ can be increased by compressive in-plane strain, while that of $Bi_2Se_3$ can be increased with tensile in-plane strain. The power factor of $n$-doped $Bi_2Se_3$ can be increased with compressive uniaxial strain, while that of $n$-doped $Bi_2Te_3$ can be increased by compressive in-plane strain. Finally, we have compared the properties of structures optimized using different functionals, and found that taking into account vdW interactions is crucial for the predictions of electronic and thermoelectric properties of strained structures, while spin-orbit interactions are less important for structure determination. In contrast, calculations on phonon frequencies suggest that spin-orbit coupling is important for accurate predictions of frequencies, while it is unclear if vdW methods, as currently implemented, contribute to accurate predictions of frequencies.

ACKNOWLEDGMENTS

This work was supported by the IHPC Independent Investigatorship ($I^3$) program. The authors acknowledge the support from the A*STAR Computational Resource Center and early discussions with O. Yazyev. The authors also thank C.K. Gan for the phonon code (force constant approach).

Table 1. Structural parameters for $Bi_2Se_3$ and $Bi_2Te_3$. Within the rhombohedral representation, the unit cell is described by the lattice constant $a_0$ and angel $\alpha$, the atom $Se_2(Te_2)$ is set to be the origin (0, 0, 0), the two Bi atoms occupy sites $\pm(\mu, \mu, \mu)$, while the two Se1(Te1) atoms are located at $\pm(\upsilon, \upsilon, \upsilon)$. The corresponding hexagonal cell edges is denoted by $a_0$' and $c_0$', and the equilibrium inter-QL distance ($d_{eqm}$) in hexagonal cell is also shown. (See Fig.1 for notation)

| | | Exp.1[b] | Exp.2[c] | LDA | LDA+SOC | PBE | PBE+SOC | revPBE | PBE-D | PBE-D+SOC | vdW-DF$^{C09}$x | vdW-DF$^{revPB}$E$_x$ |
|---|---|---|---|---|---|---|---|---|---|---|---|---|
| $Bi_2Se_3$ | $a_0$(Å) | 9.8405 | 9.841 | 9.616 | 9.552 | 10.889 | 10.617 | 12.458 | 9.938 | 9.8774 | 9.8771 | 11.18 |
| | $\alpha(°)$ | 24.304 | 24.273 | 24.593 | 24.8507 | 22.123 | 22.792 | 19.493 | 23.905 | 24.152 | 24.105 | 21.895 |
| | μ (Bi) | 0.4008 | 0.399 | 0.4009 | 0.4016 | 0.3946 | 0.3959 | 0.3868 | 0.3992 | 0.3998 | 0.4001 | 0.3946 |
| | υ (Se) | 0.2117 | 0.206 | 0.2097 | 0.2081 | 0.2228 | 0.2199 | 0.2369 | 0.2127 | 0.2113 | 0.2115 | 0.2235 |
| | $a_0$'(Å) | 4.143 | 4.138 | 4.096 | 4.111 | 4.178 | 4.195 | 4.218 | 4.116 | 4.133 | 4.125 | 4.246 |
| | $c_0$'(Å) | 28.636 | 28.64 | 27.96 | 27.758 | 31.86 | 31.01 | 36.65 | 28.95 | 28.755 | 28.757 | 32.72 |
| | $d_{eqm}$(Å) | 2.579 | 2.253 | 2.406 | 2.298 | 3.574 | 3.302 | 5.145 | 2.668 | 2.570 | 2.580 | 3.722 |
| | 2QL $d_{eqm}^a$(Å) | 2.579 | 2.253 | 2.380 | 2.28 | 3.58 | 3.38 | >4.48 | 2.68 | 2.68 | 2.48 | 3.58 |
| $Bi_2Te_3$ | $a_0$ (Å) | 10.476 | 10.473 | 10.175 | 10.124 | 10.829 | 10.628 | 12.399 | 10.664 | 10.565 | 10.368 | 11.138 |
| | $\alpha(°)$ | 24.166 | 24.167 | 24.574 | 24.806 | 23.647 | 24.223 | 20.678 | 23.305 | 23.703 | 24.277 | 23.449 |
| | μ (Bi) | 0.4000 | 0.400 | 0.4004 | 0.4012 | 0.3979 | 0.3996 | 0.3900 | 0.3985 | 0.3990 | 0.3999 | 0.3972 |
| | υ (Te) | 0.2095 | 0.212 | 0.2083 | 0.2063 | 0.2146 | 0.2109 | 0.2294 | 0.2120 | 0.2099 | 0.2094 | 0.2168 |
| | $a_0$'(Å) | 4.386 | 4.384 | 4.331 | 4.349 | 4.438 | 4.460 | 4.450 | 4.308 | 4.340 | 4.360 | 4.527 |
| | $c_0$' (Å) | 30.497 | 30.487 | 29.589 | 29.423 | 31.565 | 30.934 | 36.390 | 31.109 | 30.792 | 30.174 | 32.48 |
| | $d_{eqm}$(Å) | 2.612 | 2.764 | 2.466 | 2.331 | 3.025 | 2.736 | 4.566 | 2.822 | 2.651 | 2.582 | 3.259 |
| | 2QL $d_{eqm}^a$(Å) | 2.612 | 2.764 | 2.41 | 2.31 | 3.11 | 2.81 | >4.31 | 2.71 | 2.71 | 2.51 | 3.41 |

[a] $d_{eqm}$ for 2QL is obtained from Fig.2.

[b] From Ref. 58 (Nakajima)

[c] From Ref. 59 (Wyckoff)

Table 2. The calculated bulk moduli $B$ (GPa) of $Bi_2Te_3$ and $Bi_2Se_3$.

| | vdW-$DF^{C09}x$ | PBE-D | PBE-D+SOC | LDA | LDA+SOC | PBE+SOC | Exp. |
|---|---|---|---|---|---|---|---|
| $Bi_2Se_3$ | 42.8 | 42.92 | 41.0 | 46.68 | 48.82 | 49.45 | $32.98^a$, $53^b$ |
| $Bi_2Te_3$ | 40.2 | 35.61 | 32.5 | 42.46 | 46.56 | 42.55 | $39.47^c$, $32.5^d$ |

[a]From Ref. 81

[b]From Ref. 67

[c] From Ref. 82

[d] From Ref. 83

Table 3. Calculated zone center phonon frequencies for $Bi_2Se_3$ and $Bi_2Te_3$ bulks with and without spin-orbit coupling (SOC), the unit of frequency is in wavenumber $cm^{-1}$. The Raman and Infrared active modes are denoted with R or I.

| | | LDA+SOC | LDA | PBE+SOC[a] | PBE[a] | PBE-D+SOC | PBE-D | vdW-DF$^{C09}$x | Exp. |
|---|---|---|---|---|---|---|---|---|---|
| | $E_g^1$(R) | 42.80 | 44.272 | 38.893 | 42.128 | 38.83 | 43.92 | 43.048 | 37[b] |
| | $A_{1g}^1$(R) | 75.50 | 74.978 | 63.843 | 74.584 | 71.58 | 74.94 | 72.639 | 72.2[c] |
| | $E_u^1$(I) | 82.46 | 86.966 | 64.677 | 85.057 | 83.92 | 89.33 | 86.288 | |
| | $E_u^2$(I) | 131.06 | 136.108 | 126.819 | 132.957 | 128.80 | 135.23 | 133.62 | 131.4[c] |
| $Bi_2Se_3$ | $E_g^2$(R) | 137.99 | 142.911 | 123.984 | 138.894 | 130.47 | 138.39 | 138.172 | |
| | $A_{2u}^1$(I) | 137.44 | 145.20 | 136.692 | 142.730 | 146.67 | 153.36 | 146.575 | |
| | $A_{2u}^2$(I) | 162.89 | 171.504 | 155.439 | 167.113 | 165.87 | 169.95 | 167.856 | |
| | $A_{1g}^2$(R) | 174.46 | 180.546 | 166.346 | 179.455 | 175.47 | 182.80 | 179.762 | 174[c] |
| | $E_g^1$(R) | 42.04 | 43.34 | 35.457 | 36.358 | 38.51 | 41.84 | 40.737 | 34.356[d] |
| | $A_{1g}^1$(R) | 62.64 | 65.6 | 53.869 | 53.903 | 59.99 | 61.54 | 62.184 | 62.042[d] |
| | $E_u^1$(I) | 64.28 | 69.4 | 48.399 | 63.142 | 66.77 | 76.18 | 68.945 | |
| | $E_u^2$(I) | 94.92 | 99.65 | 91.228 | 97.399 | 95.46 | 102.68 | 100.393 | |
| $Bi_2Te_3$ | $E_g^2$(R) | 104.67 | 112.72 | 95.931 | 104.404 | 102.75 | 111.19 | 109.670 | 101.735[d] |
| | $A_{2u}^1$(I) | 96.8 | 104.03 | 95.064 | 102.569 | 104.30 | 111.89 | 103.956 | |
| | $A_{2u}^2$(I) | 120.5 | 129.47 | 118.613 | 128.220 | 126.81 | 137.21 | 128.644 | |
| | $A_{1g}^2$(R) | 131.91 | 140.38 | 127.219 | 137.2266 | 136.45 | 145.83 | 139.788 | 134.091[d] |

[a]PBE+SOC and PBE calculated data cited from Ref. 71.

[b]Ref. 84

[c]Ref. 67

[d]Ref. 85

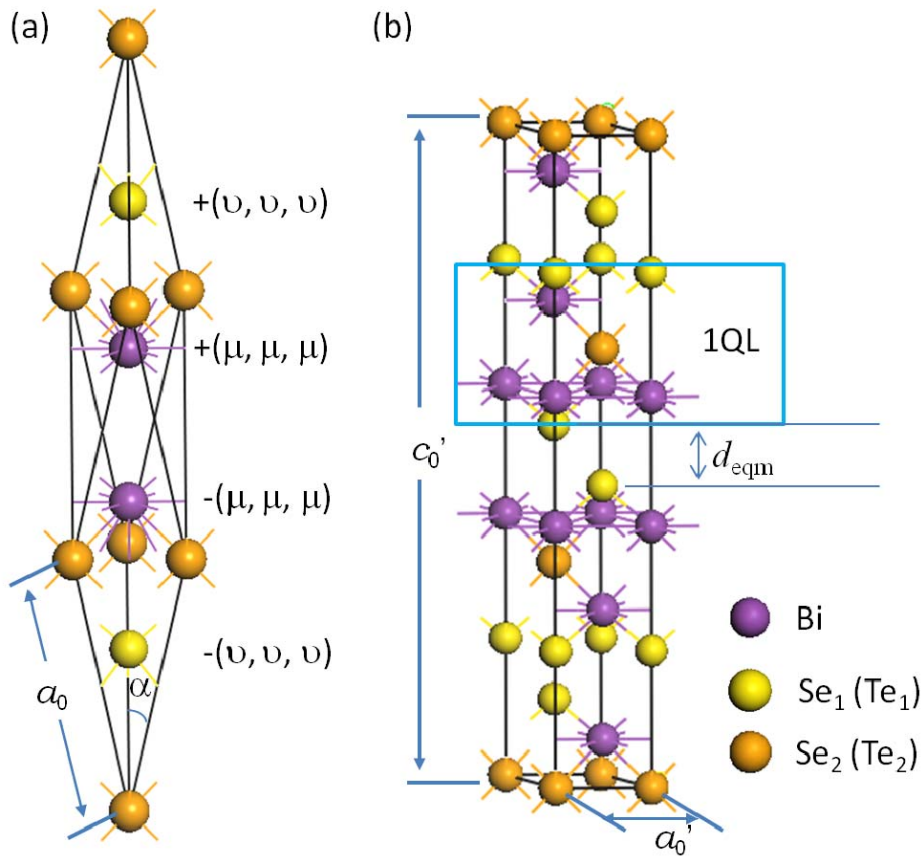

Fig. 1. Atomic structures of bulk $Bi_2Te_3$ and $Bi_2Se_3$. 1QL contains five atoms in $Se_1(Te_1)$-Bi-$Se_2(Te_2)$-Bi-$Se_1(Te_1)$ series. (a) Rhombohedral unit cell. (b) Hexagonal unit cell (containing 3QLs)

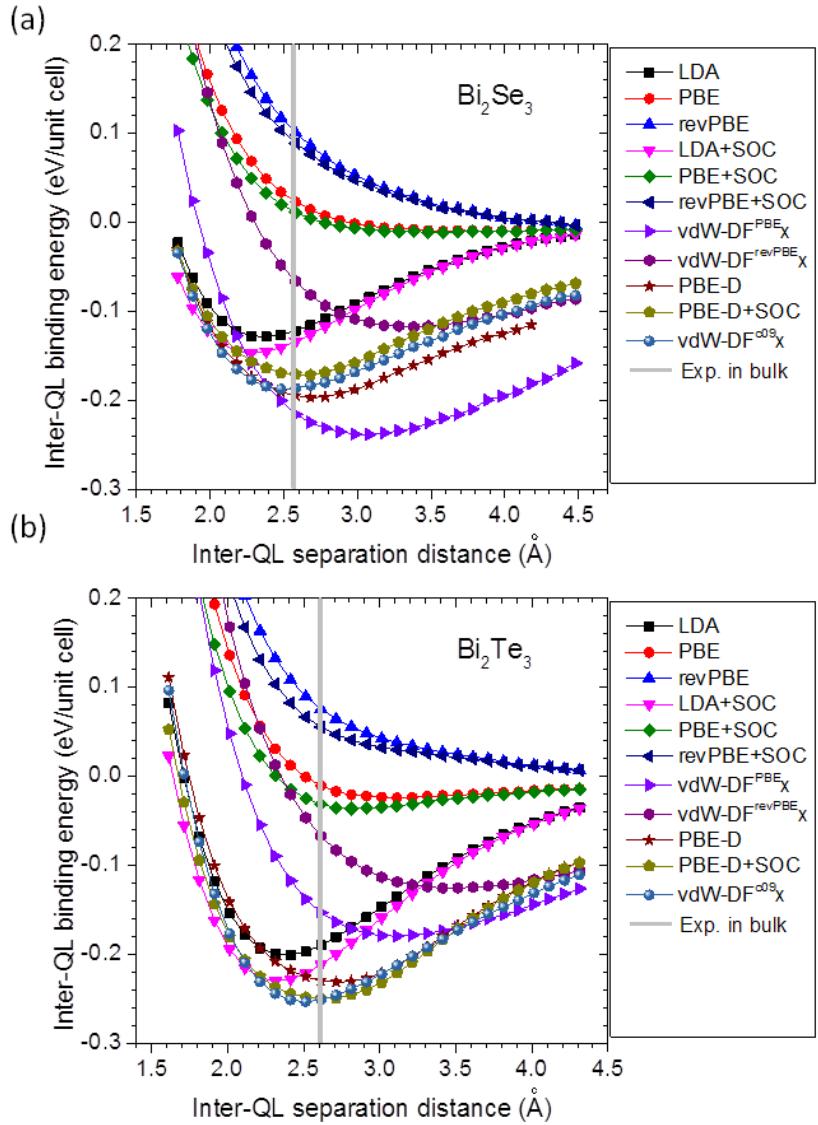

Fig. 2. Inter-QL binding energy as a function of inter-quintuple layer separation distance for (a) Bi$_2$Se$_3$ and (b) Bi$_2$Te$_3$ 2QL thin films. The inter-QL energy is computed by taking the difference between the total energy of the 2QL film, and twice that of a 1QL film in the same unit cell. Here, the internal atoms within each QL are fixed to their experimental atomic positions from the bulk. The grey line shows the value of experimental distance in bulk, which is from Ref. 58.

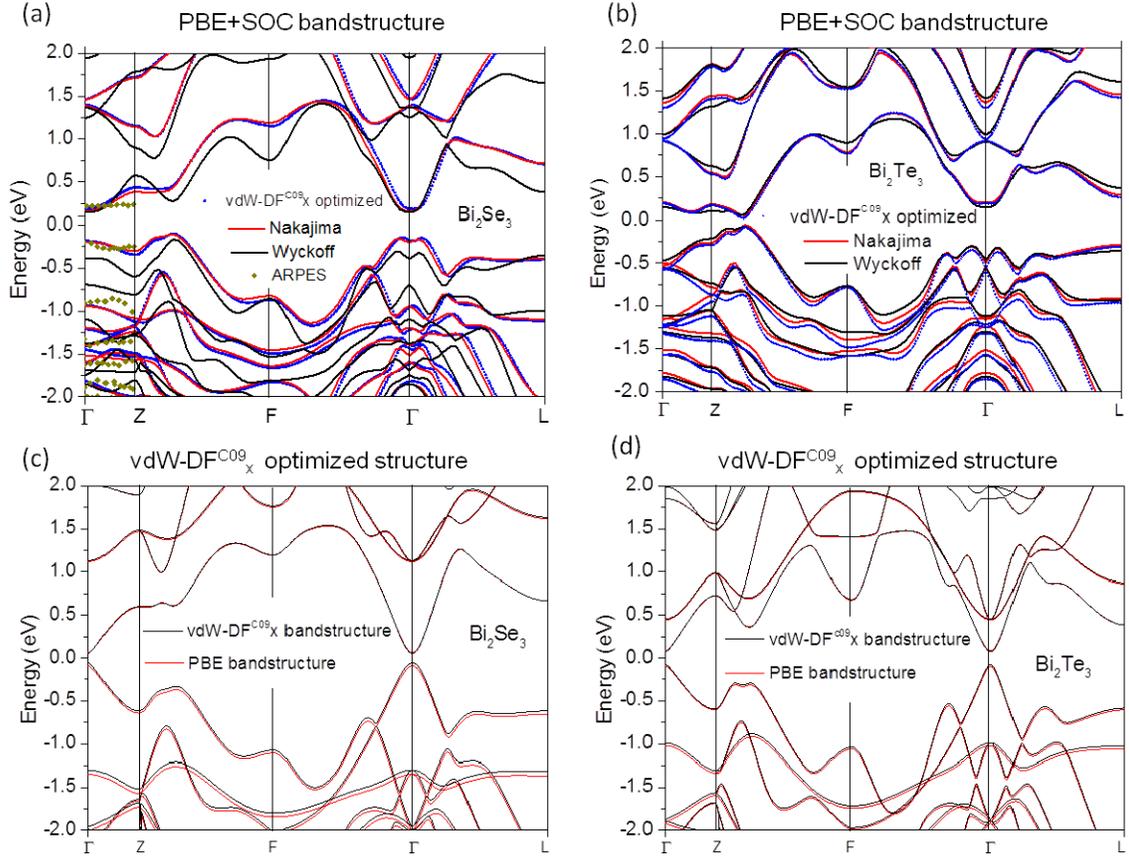

Fig. 3. Band structures of bulk (a) $Bi_2Se_3$ and (b) $Bi_2Te_3$, computed using the GGA-PBE functional with spin-orbit coupling (SOC). The experimental ARPES data for $Bi_2Se_3$ (Ref. 61) is shown along the Γ and Z high symmetry direction. The bandstructures of (c) $Bi_2Se_3$ and (d) $Bi_2Te_3$ without SOC are also calculated with vdW-DF$^{C09}_X$ and PBE functionals.

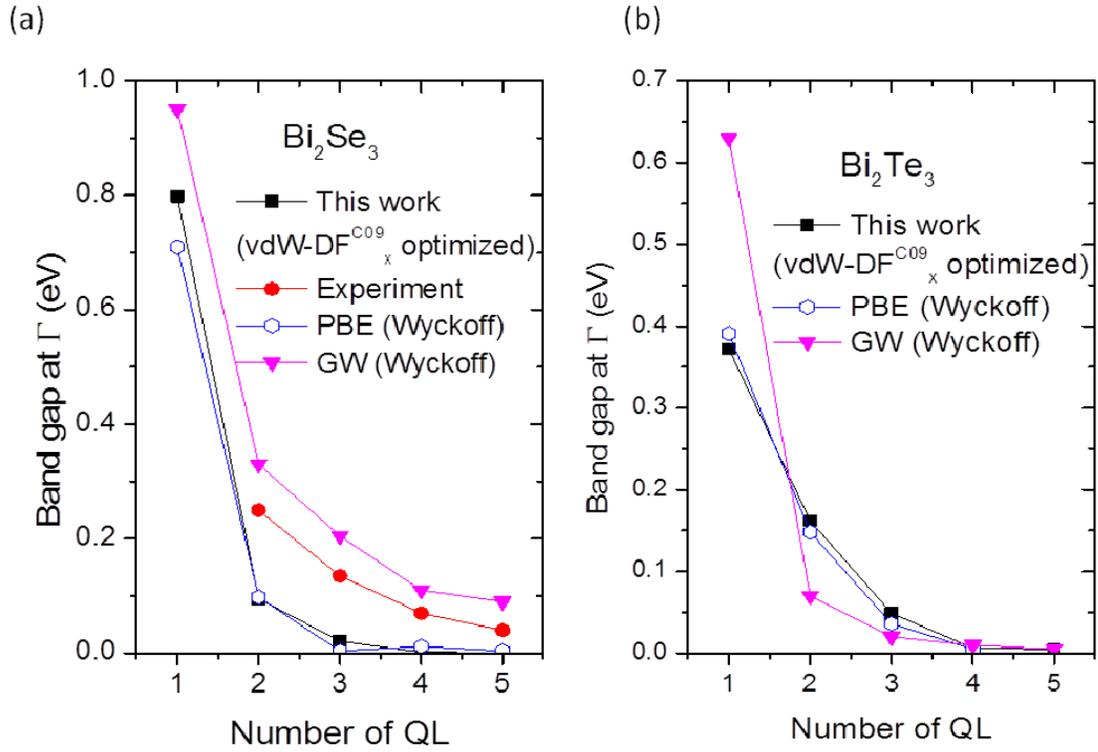

Fig. 4. Thickness dependent band gaps at the Γ point induced by the interaction between the surface states of thin film for (a) $Bi_2Se_3$ and (b) $Bi_2Te_3$. Experimental results are reproduced from Ref. 66, PBE data is obtained from Ref. 65 and the GW calculation data is derived from Ref. 64. The slab structures in PBE and GW calculations are based on the Wyckoff structures.

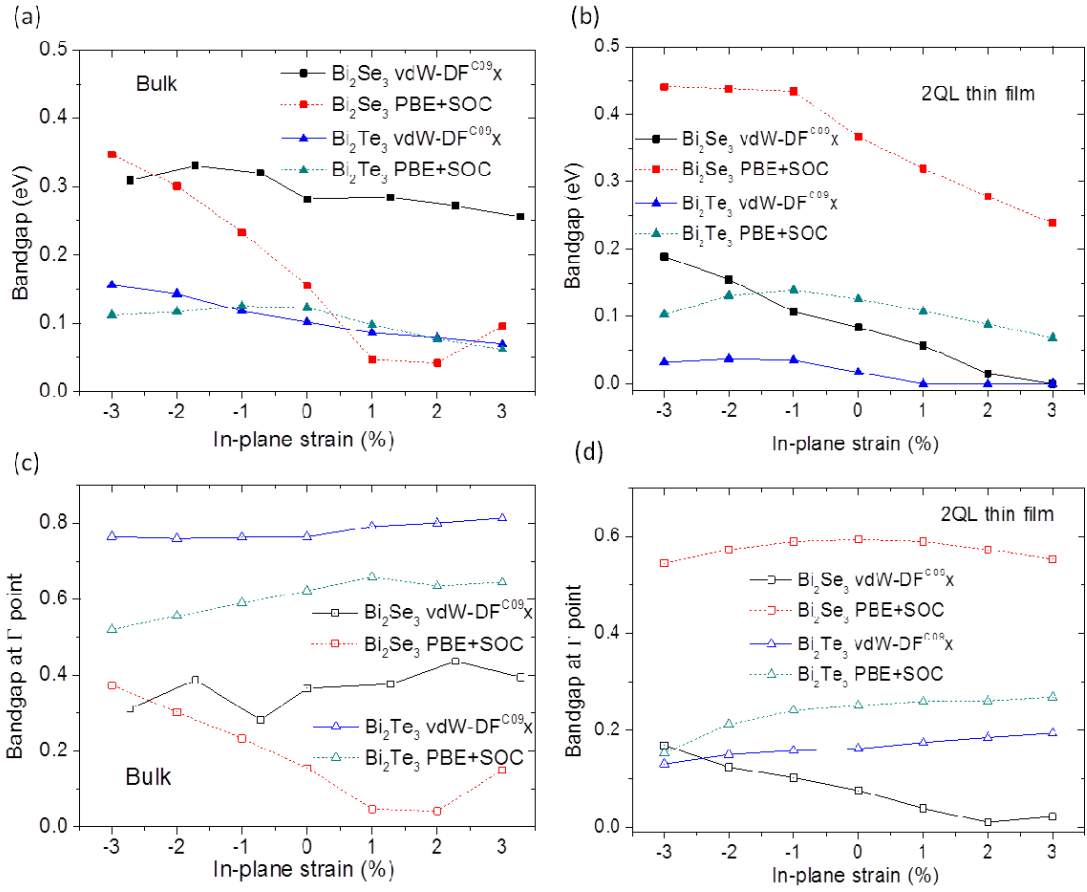

Fig. 5. (a-b) Calculated band gap as a function of in-plane strain applied to $Bi_2Se_3$ and $Bi_2Te_3$ (a) bulk materials and (b) 2QL thin films. (c-d) Band gap at $\Gamma$ point as a function of in-plane strain for (c) bulk and (d) 2QL thin films. All band structures are calculated by PBE+SOC; the notation behind each material denotes which functional is used to relax the structure.

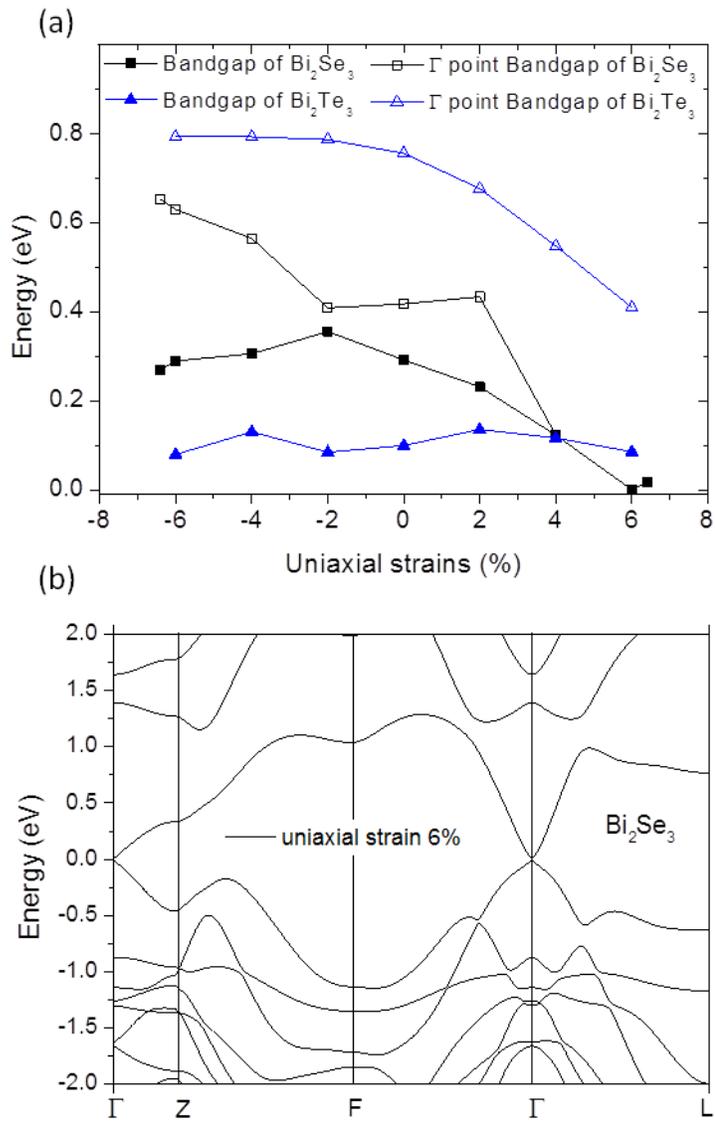

Fig. 6. (a) Direct band gap at Γ point as a function of uniaxial strain in bulk Bi$_2$Se$_3$ and Bi$_2$Te$_3$. (b) the topological phase transition at 6% uniaxial stain for Bi$_2$Se$_3$.

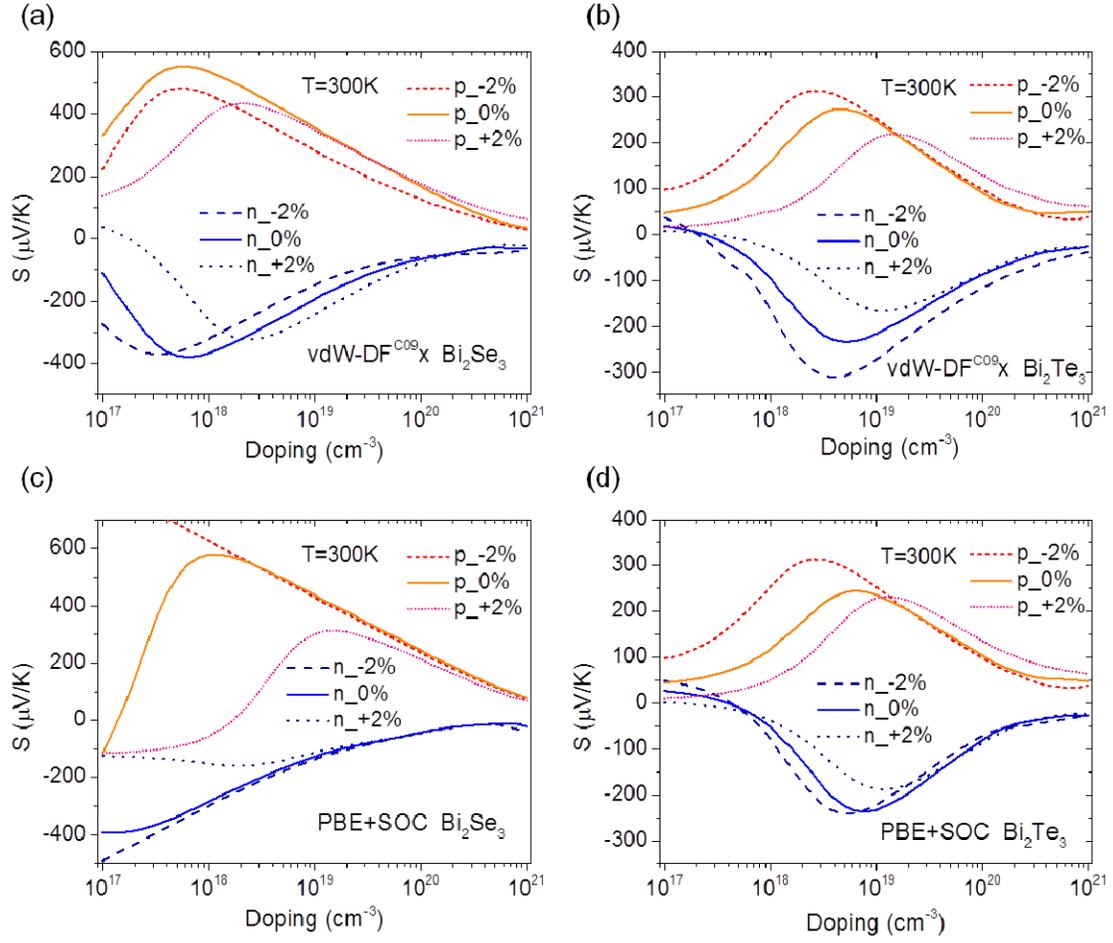

Fig. 7. (a-b) The Seebeck coefficient $S$ of vdW-DF$^{C09}$x-optimized, bulk (a) $Bi_2Te_3$ and (b) $Bi_2Se_3$ as a function of dopant concentrations, with different in-plane strains. (c-d) For comparison, the results based on the PBE+SOC relaxed structure are shown in (c) and (d) for $Bi_2Te_3$ and $Bi_2Se_3$, respectively. Please note that the scales are different for the different panels.

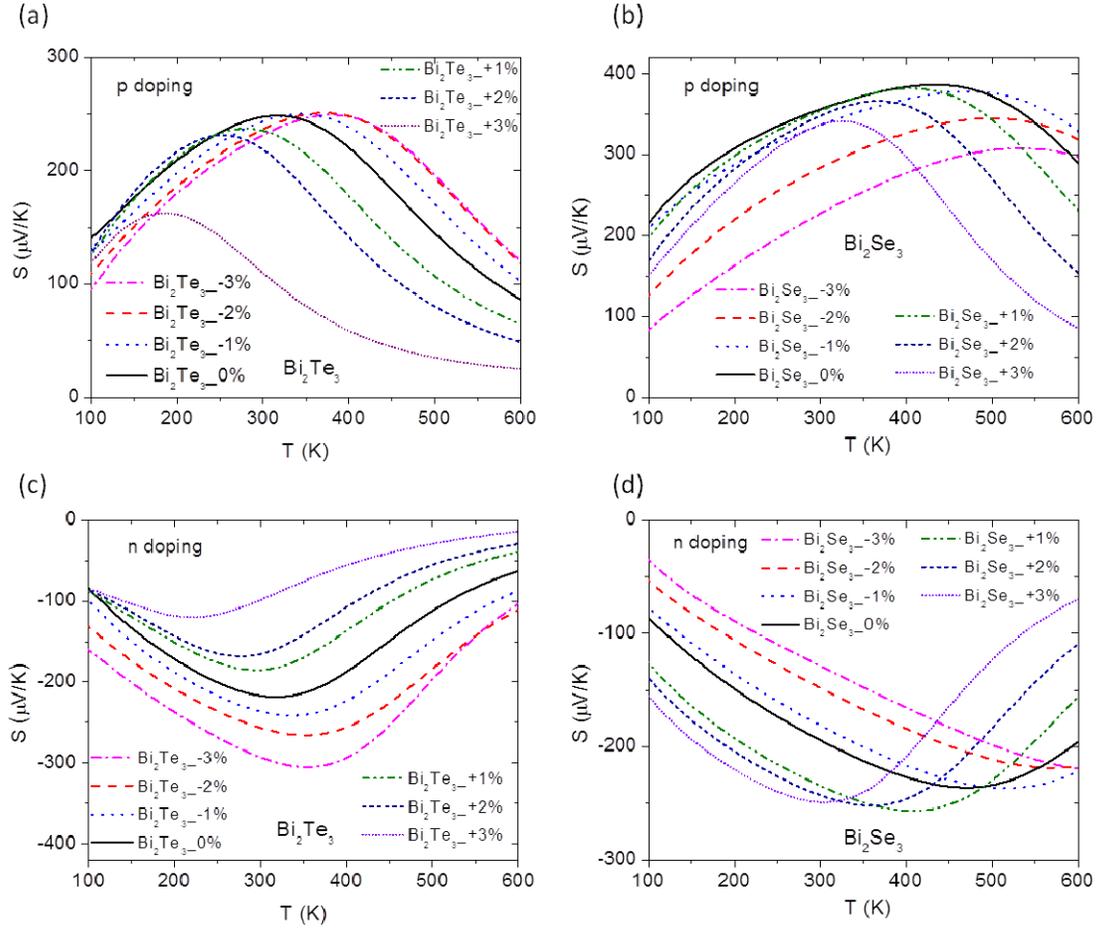

Fig. 8. Seebeck coefficient $S$ of $Bi_2Te_3$ and $Bi_2Se_3$ under various strained states, both *p*-type doping and *n*-type doping are shown. The doping level is fixed to $10^{19} cm^{-3}$. Please note that the scales are different for different panels.

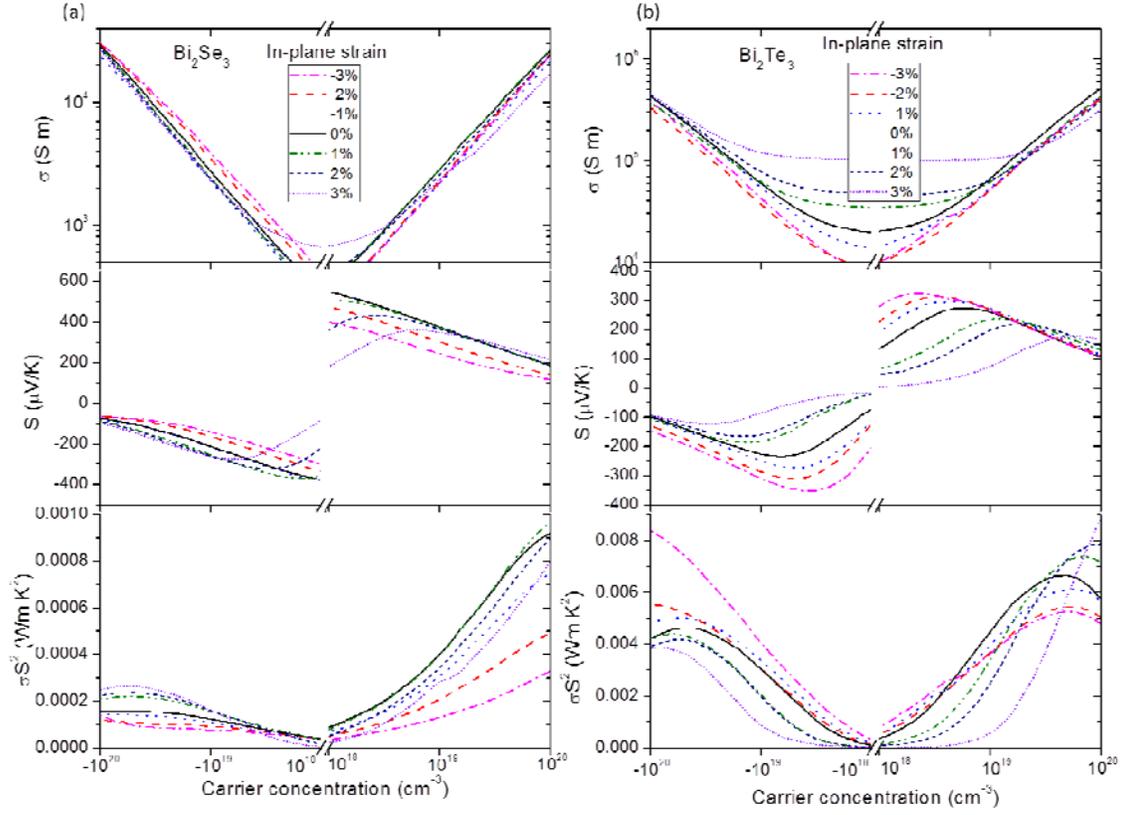

Fig. 9. The conductivity, Seebeck coefficient and power factor of (a) $Bi_2Se_3$ and (b) $Bi_2Te_3$ under different in-plane strains. Negative carrier concentrations denote electron doping. The power factor of $Bi_2Se_3$ is an order of magnitude smaller than that of $Bi_2Te_3$, because of its much smaller conductivity.

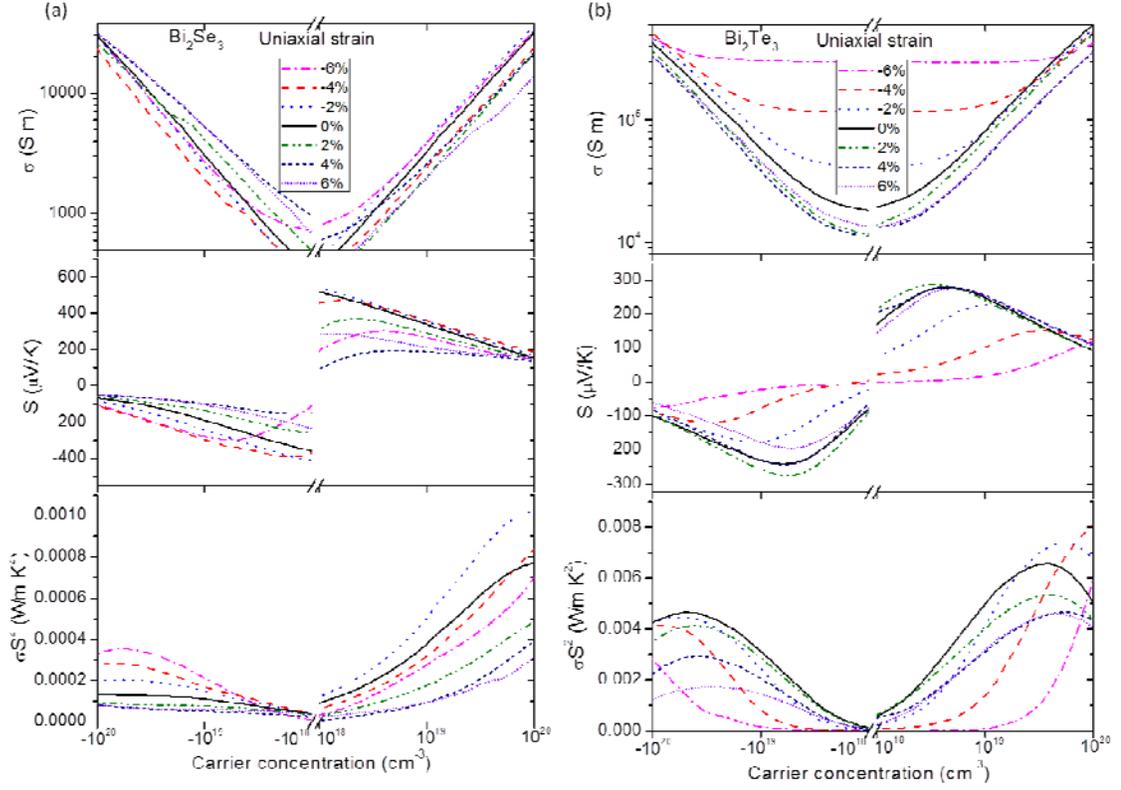

Fig. 10. The conductivity, Seebeck coefficient and power factor of (a) $Bi_2Se_3$ and (b) $Bi_2Te_3$ under different uniaxial strains. Negative carrier concentrations denote electron doping.

References:

1. D. Kraemer, B. Poudel, H. P. Feng, J. C. Caylor, B. Yu, X. Yan, Y. Ma, X. Wang, D. Wang and A. Muto, Nature materials **10**, 532 (2011).
2. I. Chowdhury, R. Prasher, K. Lofgreen, G. Chrysler, S. Narasimhan, R. Mahajan, D. Koester, R. Alley and R. Venkatasubramanian, Nature nanotechnology **4**, 235 (2009).
3. R. Venkatasubramanian, E. Siivola, T. Colpitts and B. O'Quinn, Nature **413**, 597 (2001).
4. L. Fu, C. L. Kane and E. J. Mele, Physical Review Letters **98**, 106803 (2007).
5. Y. Xia, D. Qian, D. Hsieh, L. Wray, A. Pal, H. Lin, A. Bansil, D. Grauer, Y. S. Hor, R. J. Cava and M. Z. Hasan, Nature Physics **5**, 398 (2009).
6. Y. L. Chen, J. G. Analytis, J. H. Chu, Z. K. Liu, S. K. Mo, X. L. Qi, H. J. Zhang, D. H. Lu, X. Dai and Z. Fang, Science **325**, 178 (2009).
7. H. Zhang, C. X. Liu, X. L. Qi, X. Dai, Z. Fang and S. C. Zhang, Nature Physics **5**, 438 (2009).
8. M. Z. Hasan and C. L. Kane, Reviews of Modern Physics **82**, 3045 (2010).
9. J. E. Moore, Nature **464**, 194 (2010).
10. X.-L. Qi and S.-C. Zhang, Reviews of Modern Physics **83**, 1057 (2011).
11. M. Bianchi, R. C. Hatch, J. Mi, B. B. Iversen and P. Hofmann, Physical Review Letters **107**, 086802 (2011).
12. H. M. Benia, C. T. Lin, K. Kern and C. R. Ast, Physical Review Letters **107**, 177602 (2011).
13. O. V. Yazyev, J. E. Moore and S. G. Louie, Physical Review Letters **105**, 266806 (2010).
14. M. Z. Hasan and J. E. Moore, Annual Review of Condensed Matter Physics **2**, 55 (2011).
15. P. Ghaemi, R. S. K. Mong and J. E. Moore, Physical Review Letters **105**, 166603 (2010).
16. T. Sato and H. Nakai, The Journal of chemical physics **131**, 224104 (2009).
17. E. R. Johnson, I. D. Mackie and G. A. DiLabio, Journal of Physical Organic Chemistry **22**, 1127 (2009).
18. S. Grimme, J. Antony, S. Ehrlich and H. Krieg, The Journal of chemical physics **132**, 154104 (2010).
19. M. Dion, H. Rydberg, E. Schroder, D. C. Langreth and B. I. Lundqvist, Physical Review Letters **92**, 246401 (2004).
20. T. Thonhauser, V. R. Cooper, S. Li, A. Puzder, P. Hyldgaard and D. C. Langreth, Physical Review B **76**, 125112 (2007).
21. Y. Zhang and W. Yang, Physical Review Letters **80**, 890 (1998).
22. D. Langreth, B. I. Lundqvist, S. D. Chakarova-Käck, V. Cooper, M. Dion, P. Hyldgaard, A. Kelkkanen, J. Kleis, L. Kong and S. Li, Journal of Physics: Condensed Matter **21**, 084203 (2009).
23. V. R. Cooper, Physical Review B **81**, 161104 (2010).
24. S. K. Mishra, S. Satpathy and O. Jepsen, Journal of Physics: Condensed Matter **9**, 461 (1997).
25. P. Larson, S. D. Mahanti and M. G. Kanatzidis, Physical Review B **61**, 8162 (2000).
26. P. Larson, V. A. Greanya, W. C. Tonjes, R. Liu, S. D. Mahanti and C. G. Olson, Physical Review B **65**, 085108 (2002).
27. G. Wang and T. Cagin, Physical Review B **76**, 075201 (2007).
28. S. M. Young, S. Chowdhury, E. J. Walter, E. J. Mele, C. L. Kane and A. M. Rappe, Physical Review B **84**, 085106 (2011).
29. N. F. Hinsche, B. Y. Yavorsky, I. Mertig and P. Zahn, Physical Review B **84**, 165214 (2011).
30. W. Zhang, R. Yu, H. J. Zhang, X. Dai and Z. Fang, New Journal of Physics **12**, 065013 (2010).
31. Y. Zhao, Y. Hu, L. Liu, Y. Zhu and H. Guo, Nano Letters **11**, 2088 (2011).
32. H. Chen, W. Zhu, D. Xiao and Z. Zhang, Physical Review Letters **107**, 056804 (2011).
33. M. Ye, S. V. Eremeev, K. Kuroda, M. Nakatake, S. Kim, Y. Yamada, E. E. Krasovskii, E. V. Chulkov, M. Arita and H. Miyahara, Arxiv preprint arXiv:1112.5869 (2011).

34. C. Brune, C. X. Liu, E. G. Novik, E. M. Hankiewicz, H. Buhmann, Y. L. Chen, X. L. Qi, Z. X. Shen, S. C. Zhang and L. W. Molenkamp, Physical Review Letters **106**, 126803 (2011).
35. J. L. Zhang, S. J. Zhang, H. M. Weng, W. Zhang, L. X. Yang, Q. Q. Liu, S. M. Feng, X. C. Wang, R. C. Yu and L. Z. Cao, Proceedings of the National Academy of Sciences **108**, 24 (2011).
36. T. Thonhauser, T. J. Scheidemantel, J. O. Sofo, J. V. Badding and G. D. Mahan, Physical Review B **68**, 085201 (2003).
37. T. Thonhauser, G. S. Jeon, G. Mahan and J. Sofo, Physical Review B **68**, 205207 (2003).
38. T. Thonhauser, Solid state communications **129**, 249 (2004).
39. M. T. Alam, M. P. Manoharan, M. A. Haque, C. Muratore and A. Voevodin, Journal of Micromechanics and Microengineering **22**, 045001 (2012).
40. X. Li, K. Maute, M. L. Dunn and R. Yang, Physical Review B **81**, 245318 (2010).
41. A. D. Corso and A. M. Conte, Physical Review B **71**, 115106 (2005).
42. M. V. Fischetti, Z. Ren, P. M. Solomon, M. Yang and K. Rim, Journal of Applied Physics **94**, 1079 (2003).
43. J. P. Perdew, K. Burke and M. Ernzerhof, Physical Review Letters **77**, 3865 (1996).
44. T. J. Scheidemantel, C. Ambrosch-Draxl, T. Thonhauser, J. V. Badding and J. O. Sofo, Physical Review B **68**, 125210 (2003).
45. G. K. H. Madsen and D. J. Singh, Computer Physics Communications **175**, 67 (2006).
46. D. Parker and D. J. Singh, Physical Review X **1**, 021005 (2011).
47. P. Blaha, K. Schwarz, P. Sorantin and S. B. Trickey, Computer Physics Communications **59**, 399 (1990).
48. B. R. Nag, *Electron transport in compound semiconductors*. (Springer-Verlag Berlin, 1980).
49. A. H. MacDonald, W. E. Picket and D. D. Koelling, Journal of Physics C: Solid State Physics **13**, 2675 (1980).
50. J. P. Perdew and Y. Wang, Physical Review B **45**, 13244 (1992).
51. S. Grimme, Journal of computational chemistry **27**, 1787 (2006).
52. V. Barone, M. Casarin, D. Forrer, M. Pavone, M. Sambi and A. Vittadini, Journal of computational chemistry **30**, 934 (2009).
53. G. Román-Pérez and J. M. Soler, Physical Review Letters **103**, 096102 (2009).
54. P. H. Tan, W. P. Han, W. J. Zhao, Z. H. Wu, K. Chang, H. Wang, Y. F. Wang, N. Bonini, N. Marzari and N. Pugno, Nature Materials **11**, 294 (2012).
55. Zhao Yanyuan, Luo Xin, Li Hai, Zhang Jun, Paulo Antonio Trindade Araujo, Gan CheeKwan, W. J., Zhang Hua, Quek Su Ying, Mildred S. Dresselhaus and X. Qihua,  (2012).
56. O. A. Vydrov and T. Van Voorhis, Journal of Chemical Physics **133**, 244103 (2010).
57. T. Bjorkman, A. Gulans, A. V. Krasheninnikov and R. M. Nieminen, Physical Review Letters **108**, 235502 (2012).
58. S. Nakajima, Journal of Physics and Chemistry of Solids **24**, 479 (1963).
59. R. W. G. Wyckoff,  (New York: John Wiley & Sons, 1964).
60. S. A. Semiletov, Tr. Inst. Kristallogr., Akad. Nauk SSSR **10**, 76 (1954).
61. V. A. Greanya, W. C. Tonjes, R. Liu, C. G. Olson, D. Y. Chung and M. G. Kanatzidis, Journal of Applied Physics **92**, 6658 (2002).
62. M. S. Hybertsen and S. G. Louie, Physical Review B **34**, 5390 (1986).
63. J. Black, E. M. Conwell, L. Seigle and C. W. Spencer, Journal of Physics and Chemistry of Solids **2**, 240 (1957).
64. O. V. Yazyev, E. Kioupakis, J. E. Moore and S. G. Louie, Physical Review B **85**, 161101 (2012).
65. C. X. Liu, H. J. Zhang, B. Yan, X. L. Qi, T. Frauenheim, X. Dai, Z. Fang and S. C. Zhang, Physical Review B **81**, 041307 (2010).

66. Y. Zhang, K. He, C. Z. Chang, C. L. Song, L. L. Wang, X. Chen, J. F. Jia, Z. Fang, X. Dai and W. Y. Shan, Nature Physics **6**, 584 (2010).
67. R. Vilaplana, D. Santamaría-Pérez, O. Gomis, F. Manjón, J. González, A. Segura, A. Muñoz, P. Rodríguez-Hernández, E. Pérez-González and V. Marín-Borrás, Physical Review B **84**, 184110 (2011).
68. M. Lazzeri and F. Mauri, Physical Review Letters **90**, 36401 (2003).
69. C. K. Gan, Y. P. Feng and D. J. Srolovitz, Physical Review B **73**, 235214 (2006).
70. G. Kresse, J. Furthmüller and J. Hafner, EPL (Europhysics Letters) **32**, 729 (2007).
71. W. Cheng and S. F. Ren, Physical Review B **83**, 094301 (2011).
72. H. J. Goldsmid and R. W. Douglas, British Journal of Applied Physics **5**, 386 (1954).
73. H. J. Goldsmid, Proceedings of the Physical Society **71**, 633 (1958).
74. H. J. Goldsmid, A. R. Sheard and D. A. Wright, British Journal of Applied Physics **9**, 365 (1958).
75. T. M. Tritt, Annual Review of Materials Research **41**, 433 (2011).
76. M. S. Park, J.-H. Song, J. E. Medvedeva, M. Kim, I. G. Kim and A. J. Freeman, Physical Review B **81**, 155211 (2010).
77. B. Y. Yavorsky, N. F. Hinsche, I. Mertig and P. Zahn, Physical Review B **84**, 165208 (2011).
78. We note that the bandstructures obtained from Wien2k are slightly different from that obtained with QE. This difference may be due to the different implementations of spin-orbit coupling in the two codes.
79. Y. S. Hor, A. Richardella, P. Roushan, Y. Xia, J. G. Checkelsky, A. Yazdani, M. Z. Hasan, N. P. Ong and R. J. Cava, Physical Review B **79**, 195208 (2009).
80. N. F. Mott and E. A. Davis, *Electronic processes in non-crystalline materials*. (Clarendon, Oxford, 1979).
81. J. J. Hamlin, J. R. Jeffries, N. P. Butch, P. Syers, D. A. Zocco, S. T. Weir, Y. K. Vohra, J. Paglione and M. B. Maple, Journal of Physics: Condensed Matter **24**, 035602 (2011).
82. J. O. Jenkins, J. A. Rayne and R. W. Ure Jr, Physical Review B **5**, 3171 (1972).
83. R. Vilaplana, O. Gomis, F. J. Manjón, A. Segura, E. Pérez-González, P. Rodríguez-Hernández, A. Muñoz, J. González, V. Marín-Borrás, V. Muñoz-Sanjosé, C. Drasar and V. Kucek, Physical Review B **84**, 104112 (2011).
84. J. Zhang, Z. Peng, A. Soni, Y. Zhao, Y. Xiong, B. Peng, J. Wang, M. S. Dresselhaus and Q. Xiong, Nano letters **11**, 2407 (2011).
85. K. M. F. Shahil, M. Z. Hossain, D. Teweldebrhan and A. A. Balandin, Applied Physics Letters **96**, 153103 (2010).